# Origin of Negative Thermal Expansion and Pressure Induced Amorphization in Zirconium Tungstate from Machine-Learning Potential

Ri He[1], Hongyu Wu[1], Yi Lu[2,3], Zhicheng Zhong[1,4*]

[1]Key Laboratory of Magnetic Materials Devices & Zhejiang Province Key Laboratory of Magnetic Materials and Application Technology, Ningbo Institute of Materials Technology and Engineering, Chinese Academy of Sciences, Ningbo 315201, China

[2]National Laboratory of Solid State Microstructures and Department of Physics, Nanjing University, Nanjing 210093, China

[3]Collaborative Innovation Center of Advanced Microstructures, Nanjing University, Nanjing 210093, China

[4]China Center of Materials Science and Optoelectronics Engineering, University of Chinese Academy of Sciences, Beijing 100049, China

* zhong@nimte.ac.cn

# Abstract

Understanding various macroscopic pressure-volume-temperature properties of materials on the atomistic level has always been an ambition for physicists and material scientists. Particularly, some materials such as zirconium tungstate ($ZrW_2O_8$), exhibit multiple exotic properties including negative thermal expansion (NTE) and pressure-induced amorphization (PIA). Here, using machine-learning based deep potential, we trace both of the phenomena in $ZrW_2O_8$ back to a common atomistic origin, where the nonbridging O atoms play a critical role. We demonstrate that the nonbridging O atoms confer great flexibility to vibration of polyhedrons, and kinetically drive volume shrinking on heating, or NTE. In addition, beyond a certain critical pressure, we find that the migration of nonbridging O atoms leads to additional bond formation that lowers the potential energy, suggesting that the PIA is a potential-driven first-order phase transition. Most importantly, we identify a second critical pressure beyond which the amorphous phase of $ZrW_2O_8$ undergoes a "hidden" phase transition from a reversible phase to an irreversible one.

## Introduction

Macroscopic properties of materials are deeply rooted in their underlying atomic arrangement and interaction, and understanding these properties on an atomistic level has been a long-sought endeavor[1]. Temperature and external pressure, two of the most commonly used "tuning knobs" for manipulating physical properties of materials to date, has led to the discovery of numerous exotic phenomena such as negative thermal expansion (NTE)[2, 3, 4], liquid-liquid second critical point[5, 6, 7], and pressure-induced high-temperature superconductivity[8, 9, 10]. Establishing a clear relationship between these macroscopic properties and the atomistic behavior is not only important from a fundamental perspective, but also holds the key for rational material design. It yet remains a highly nontrivial task due to the large number of degrees of freedom inherent to a macroscopic system, which are usually difficult to be captured in microscopic theories.

Take the NTE for example, which refers to the counter-intuitive phenomenon that certain materials contract upon heating[3, 4]. NTE materials are relatively rare and $ZrW_2O_8$ is the one of the most representative ones among them. $ZrW_2O_8$ exhibits volume reduction with increasing temperature from 0.3 K to 1050 K[11]. The nature of the local mechanism responsible for its NTE remains controversial as the exact atomic structure details on heating can only be indirectly inferred from experiment with certain model assumption[12, 13, 14]. One of the most common explanations is the Rigid Unit Modes (RUM) model, in which the transverse vibrations of O atoms can only occur through coupled liberations of tetrahedra and octahedra forming the structure[15, 16, 17], although its validity needs to be verified by microscopic calculations that has so far remained elusive.

In addition to NTE, $ZrW_2O_8$ exhibits another interesting property, namely the pressure-induced amorphization transition (PIA). Perottoni *et al.* showed that above 1.5 GPa, crystalline $ZrW_2O_8$ transforms to an amorphous phase that persists after pressure release at ambient conditions[2, 18, 19]. While some theoretical research suggests that the PIA and NTE may have a common origin[20], the exact mechanism of PIA in $ZrW_2O_8$ are still heavily debated[21, 22, 23, 24, 25], as the complexity of disordered atomic structure poses serious challenges to traditional theoretical methods such as the classical atomic force field, which describes bonded and nonbonded interactions in solids using some prescribed functions with fitted parameters that severely limit the flexibility of the force fields[14, 15]. The lack of periodicity, and consequently a small enough unit cell, in amorphous materials also prevents its accurate description using methods such as density functional theory (DFT), whose computation complexity scales polynomially with the unit cell size.

To solve this outstanding problem and determine the atomistic origin of the NTE and PIA, in this study, we develop a machine learning (ML)-based deep potential (DP) for $ZrW_2O_8$ using training data from DFT calculations. ML-based potentials have the flexibility and non-linearity necessary to describe complex ordered and disordered interatomic environment from the perspective of "many-body" using a deep neural network[26, 27]. While still in infancy state, they have been successfully applied to the investigation of various process in complex material systems[28], including phase transition[29, 30, 31], nucleation[32], chemical catalysis[33], gas combustion[34], and atomic layer deposition of film growth[35]. Herein, we find that the well-trained DP can describe the structure and other structure-related properties of $ZrW_2O_8$ over a wide temperature and pressure range. Using the DP, the calculated NTE and PIA in $ZrW_2O_8$ show great agreement with experimental results. Importantly, DP calculated results reveal that the NTE and PIA originate from a common factor, namely the nonbridging O atoms. For NTE, the nonbridging O confers great flexibility to the rotation of tetrahedra and octahedra that lead to volume contraction upon heating. In addition, the nonbridging O atoms move into a neighboring polyhedron above a critical pressure of 1.4 GPa and form additional bonds to lower the thermodynamical potential energy, driving a first-order amorphization phase transition as observed in experiments. Furthermore, we predict that further increasing the pressure to a second critical value of 3.8 GPa induces a "hidden" phase transition from an $a^I$ reversible phase to an $a^{II}$ irreversible one.

# Results

## Accurate Deep Learning Potential

The crystal structure of $ZrW_2O_8$ at 0 K to 430 K is cubic phase with $P2_13$ space group as illustrated in Fig. 1. The unit cell consists of a network of corner-shared octahedral $ZrO_6$ and tetrahedral $WO_4$. The six octahedral $ZrO_6$ are equivalent and each share all their six corners with neighboring $WO_4$ tetrahedra, while the eight tetrahedral $WO_4$ each share only three of their four O atoms with the neighboring octahedral $ZrO_6$, leaving one W-O bond dangling. We refer the corresponding O atom as the nonbridging O (marked by green spheres in Fig. 1). The tetrahedral $WO_4$ can be divided into two types depending on the length of nonbridging W-O bond. The $WO_4$ tetrahedrons with shorter (longer) W-O bond are labeled as $W^IO_4$ ($W^{II}O_4$). All the crystalline phase related results in the main text are based on the cubic $\alpha$-$ZrW_2O_8$, while the discussion of $\beta$-phase and $\gamma$-phase can be found in Supplementary Materials. The DFT optimized ground-state cubic phase is used as initial structure to explore all the characteristic configurations of high-temperature crystal and high-pressure amorphous. The 20460 configurations with different temperatures and pressures used in the training dataset are explore by the DP-GEN concurrent learning procedure[36], and the deep neural network-based DP is trained using the DeePMD method (for details, see Methods)[26, 27].

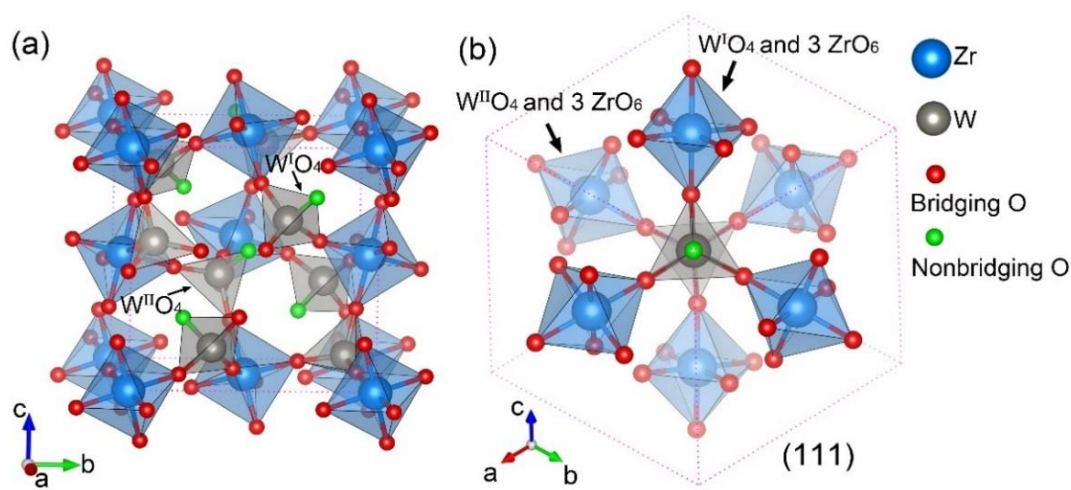

**Fig.1 Structure of cubic $ZrW_2O_8$ unit cell**. (a) The cubic cell consists of a network of corner-shared octahedral $ZrO_6$ and tetrahedral $WO_4$. The octahedral $ZrO_6$ are equivalent but the tetrahedral WO$_4$ can be divided into W$^I$O$_4$ and W$^{II}$O$_4$ depending on the length of nonbridging W-O bond. (b) <111> direction view of cubic cell.

Once the well-trained deep neural network-based DP is obtained, it is first benchmarked against DFT results to access its accuracy. We compare the energies and atomic forces calculated using the DP and DFT for 20460 configurations in the final training dataset. We find a good agreement between DP and DFT calculated energies and forces with a mean absolute error are 2.61 meV/atoms and 0.12 eV/Å, respectively, as shown in Fig. 2a, b. In comparison, the classical force field of ZrW$_2$O$_8$ proposed by Pryde et al.[15] vastly overestimates atomic forces due to the limitations imposed by the preset functions and fitted parameters (see Fig. S1), and fails to predict atomic interactions involving the bond formation and breaking under pressure[37]. The equations of state of cubic phase calculated by DFT and DP are presented in Fig. 2c. The DP model reproduces well the DFT energy profile over a wide range of lattice constants from 8.5 Å to 9.6 Å. The equilibrium lattice constants optimized by DFT and DP for cubic phase are 9.215 Å and 9.223 Å, respectively. We also use the DP to calculate the elastic constants for cubic phases and compare the values with DFT results. The results are shown in Table SI, again demonstrating great agreement between DP and DFT. Phonon dispersion spectrum is the quantum mechanical description of an atomic vibrational motion along different wave vectors, where the atomic force plays an important role. Accurate calculation of phonon dispersion is a strict criterion for testing the accuracy of DP. The calculated phonon dispersion of the cubic ZrW$_2$O$_8$ by the DP is shown in Fig. 2d. The 44-atom unit cell produces 3 acoustic and 129 optical branches. The 3 acoustic branches exist only in the low-frequency regime, and approach zero frequency towards the high-symmetry Gamma point. No negative phonon frequencies in the phonon dispersion indicates the dynamical stability of ZrW$_2$O$_8$ at room temperature and ambient pressure, which agrees well with the DFT results[38]. The systematic benchmark shows that the DP has excellent DFT-level accuracy, and is capable of predicting a range of temperature and pressure properties of ZrW$_2$O$_8$ from first principles.

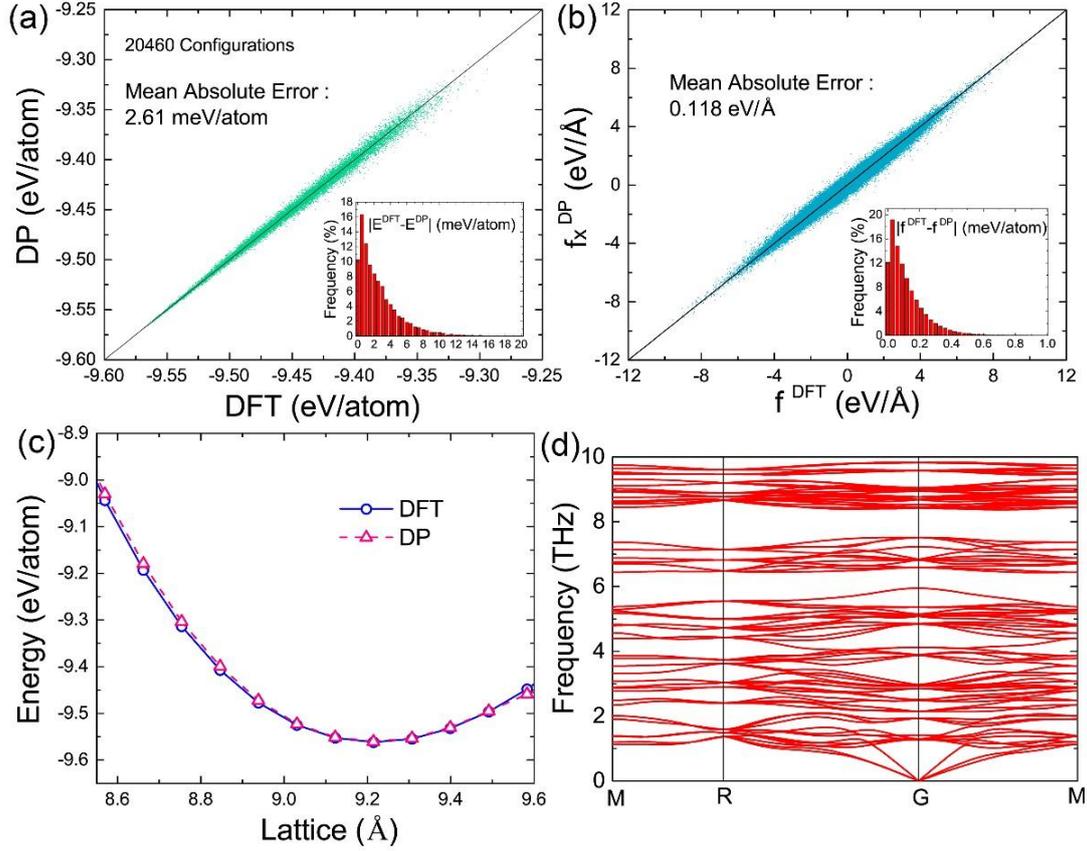

**Fig. 2 The benchmark test of DP against DFT results.** Comparison of energies (a) and atomic forces (b) calculated using the DP and DFT for all configurations. DP and DFT energies for different hydrostatic pressure (c) and phonon dispersion relations (d) for cubic phase.

**The Origin of Negative Thermal Expansion**

Once the DFT-level accuracy of DP is confirmed, we can use it to investigate the evolution of crystal structural properties of $ZrW_2O_8$ on heating by deep potential molecular dynamics (DPMD) simulations using a much larger system. Fig. 3a shows the lattice constant variation of a 7.3 nm ×7.3 nm ×7.3 nm supercell (includeing 22528 atoms) with increasing temperature at ambient pressure. It can be clearly seen that the lattice constant exhibits a continuous shrink from 0 K to 1000 K (also see Supplementary Fig. S2). This result demonstrates an excellent isotropic NTE property of $ZrW_2O_8$. The DP calculated coefficient of thermal expansion is $-6.2 \times 10^{-6}$ K$^{-1}$, in good agreement with reported experimental results of $-4$ to $-9 \times 10^{-6}$ K$^{-1}$[11, 39]. The neutron pair distribution functions $G(r)$ at room temperature obtained with the DPMD simulations is shown in Fig. 3b. The excellent agreement between the calculation and the experimental result [23] confirms that the DP can well capture the atomic structural characteristics of $ZrW_2O_8$ at room temperature.

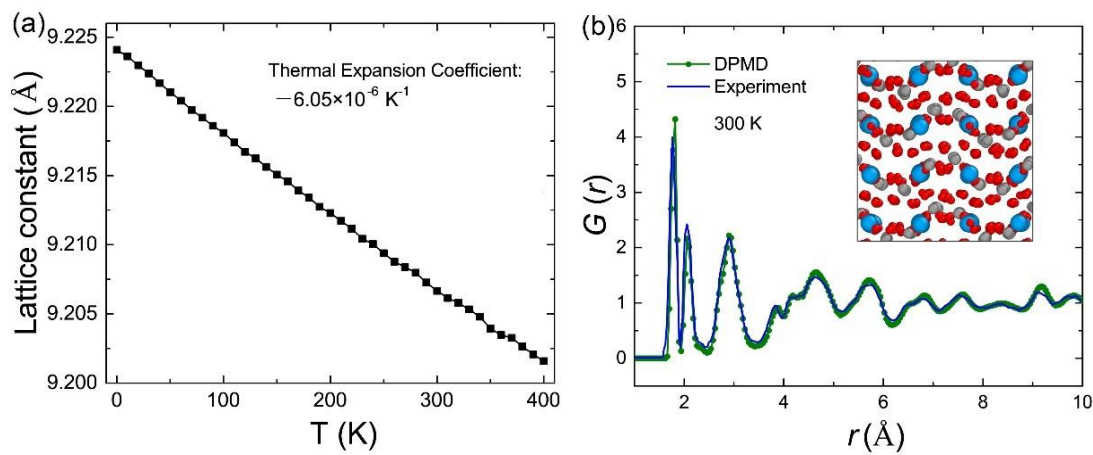

**Fig. 3 Crystal structural properties of $ZrW_2O_8$ on heating.** (a) Variation of cubic $ZrW_2O_8$ lattice parameter with temperature from DPMD at ambient pressure. (b) Neutron weighted pair distribution functions $G(r)$ from cubic $ZrW_2O_8$ at 300 K and 1 bar calculated from DPMD and experiment.

Despite the extensive study, the exact origin of NTE in $ZrW_2O_8$ is still controversial and has been attributed to a variety of mechanisms. For instance, the "RUM" model suggests that the transverse vibrations of the O atoms in the middle of W-O-Zr linkage are the result of whole-body rotations and translations of the linked octahedral $ZrO_6$ and tetrahedral $WO_4$ with essentially no distortion, pulling the Zr and W atoms closer and thus causing the NTE (a sketch is shown in Fig. S3)[11, 15, 16, 17]. In contrast, the "Tent" model assumes a stiff W-O-Zr linkage, and NTE derives from the translational motion of tetrahedral $WO_4$ along the ⟨111⟩ direction and the correlated motion of the three nearest octahedral $ZrO_6$ [12, 13]. In other words, the main controversy between the RUM and Tent models lies in the relative stiffness of the W-O-Zr linkage.

To resolve the above controversy, we investigate the thermal expansion of neighboring atom pair distances (or bond length) and the degree of rigidity of corner-sharing polyhedrons. In DPMD simulations, each atom at $\mathbf{r}_i$ vibrates about its mean position $\langle\mathbf{r}_i\rangle$. The atomic distances between two atoms can thus take an "apparent" value $\mathbf{R} = |\langle\mathbf{r}_i\rangle - \langle\mathbf{r}_j\rangle|$, which is the difference between their mean positions, and a time-dependent "true" value $r = |r_i - r_j|$ [3, 14]. The former defines the lattice constants and can be directly determined by x-ray or neutron diffraction measurements. The latter contains more detailed dynamical properties of the lattice but is not easily probed in experiments. A similar consideration can be used to define "apparent" and "true" angles of W-O-Zr linkage. The DPMD results obtained for apparent and true bond length (atom pair distance) of W-O, Zr-O, Zr-W, Zr-Zr, and angle of Zr-O-W at different temperatures are presented in Fig. 4 (for $W^IO_4$ and three nearest $ZrO_6$) and Fig. S4 (for $W^{II}O_4$ and three nearest $ZrO_6$). It clearly indicates that the W-O (Zr-O) bonds display a negative thermal expansion coefficient of $-1.542 \times 10^{-5}$ K$^{-1}$ ($-0.377 \times 10^{-5}$ K$^{-1}$) for the apparent bond length, but a positive one of 0.504 K$^{-1}$ ($1.548 \times 10^{-5}$ K$^{-1}$) for the true bond length. The DPMD results reveal that the asymmetry of atomic interaction potential causes an increase in the mean instantaneous distance between the bonded atoms as the temperature increases, but the distance between the mean positions of two atoms tend to decrease. These results show that both the RUM and Tent models are only partially right. On one hand, the observed expansion of the $WO_4$ tetrahedra and $ZrO_6$ octahedra invalidates the rigid-polyhedron assumption in RUM model. On the other hand, while the apparent W-O-Zr angle Φ remains nearly constant at ~155.5 degrees, in a seemingly agreement with the assumption of the Tent mode, the true W-O-Zr angle <φ> decrease drastically with increasing temperature as shown in Fig. 4c. In reality, the expansion of true W-O and Zr-O bonds are overcompensated by the decrease of the true W-O-Zr angle caused by the transverse vibrations of the O atom in the middle of the Zr-O-W linkage (Fig. 4f), and their combined effect results in the observed NTE, as reflected in the reduction of the overall lattice constants as well as the Zr-W and Zr-Zr distances (Fig. 4d, e and Supplementary Fig. S5).

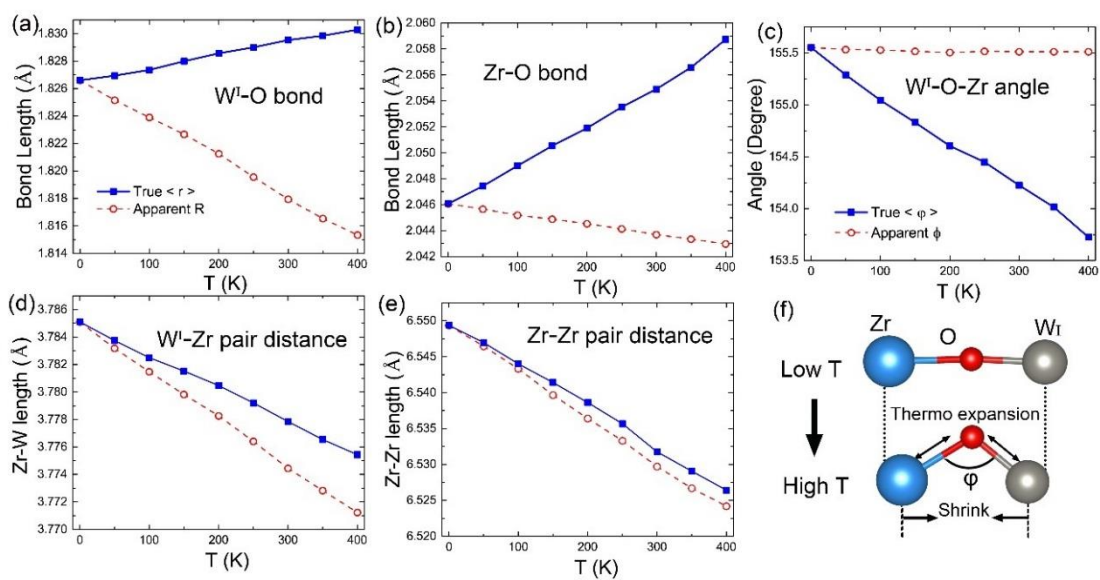

**Fig. 4 Thermal expansion of neighboring atom pair distances and degree of corner-sharing polyhedrons.** "True" and "apparent" bond length/atom pair of (a) W-O, (b) Zr-O, (d) Zr-W, (e) Zr-Zr, and angle of (c) Zr-O-W as a function of temperature. (f) The sketch of bond length and angle variation on heating.

**Pressure Induced Amorphous Phase and its Reversibility**

The amorphization transition under high pressure is another key feature of $ZrW_2O_8$. While the NTE and the PIA are believed to be closely linked [18, 23, 40], the detailed complex atomic structure of amorphous $ZrW_2O_8$ has been a subject of debate. To resolve this issue, we further perform DPMD simulations of $ZrW_2O_8$ under different pressure (0.2 GPa ~ 4.4 GPa) at 300 K for 50 ps, with subsequent pressure relaxation to 1 bar for 100 ps. The structural properties as a function of time are shown in Fig.5a, and the lattice constant dependence on pressure for compression (red line) and decompression (blue line) processes are shown in Fig. 5b. Upon application of pressure, it can be observed that the lattice constant decreases slightly by 0.05 Å from 1 bar to 1.3 GPa. Analysis of structural partial radial distribution functions (RDFs), obtained from DPMD simulations, can offer a detailed description of the observed structural phase transition from crystal to amorphous on an atomistic level. In the pressure range between 1 bar to 1.3 GPa, the partial O-O RDFs in Fig. 6a clearly show that the $ZrW_2O_8$ maintains perfect crystalline structure. The corresponding atomic structures of crystalline phase are shown in Supplementary Fig. S6 a-c. The crystalline $ZrW_2O_8$ under pressure still exhibits excellent NTE property (see Supplementary Fig. S7). Increasing the pressure beyond 1.4 GPa, the O-O RDF peaks vanish at distances larger than 8 Å (see Fig. 6a). This substantial modification indicates that $ZrW_2O_8$ undergoes an amorphization transition. The snapshots of corresponding atomic structures for 1 bar to 2.5 GPa in Supplementary Fig. S6 also clearly show that a disordered structure form above 1.4 GPa. Accompanying the PIA is a sudden decrease of lattice constant from 9.16 to 8.89 Å (Fig. 6b), which shows the first-order nature of the phase transition. The amorphous $ZrW_2O_8$ does not exhibits NTE properties as shown in Supplementary Fig. S7. The calculated PIA critical pressure 1.4 GPa is well comparable to the experimental value of 1.5 GPa [2].

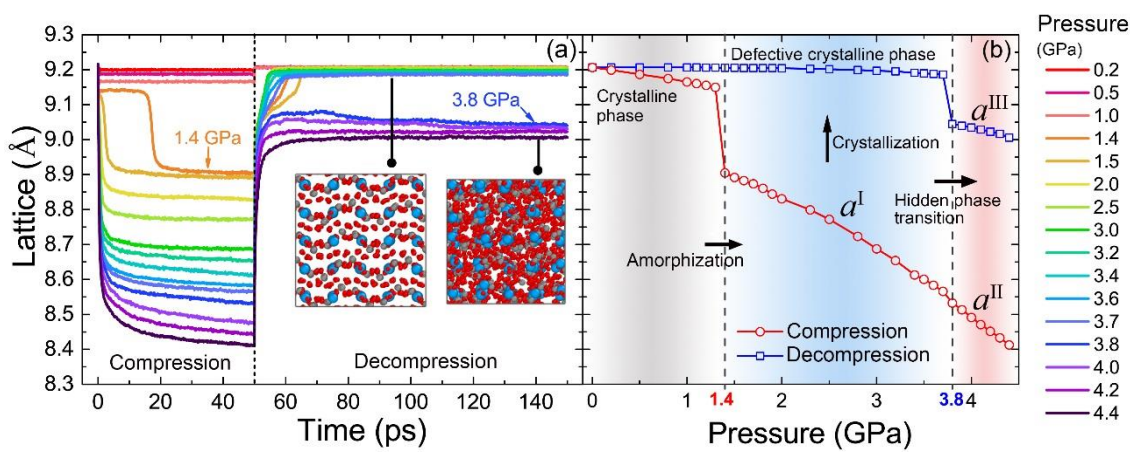

Fig. 5 **Structural properties of $ZrW_2O_8$ with compression and decompression.** (a) Variation of lattice parameter with time on compression from 0.2 GPa to 4.4 GPa at 300 K and then upon decompression. (b) The lattice parameter depends on pressure for compression after 50 ps (red line) and decompression after 100 ps (blue line).

To reveal the exact origin of PIA in $ZrW_2O_8$, we explore the energy stability of cubic and amorphous phases as a function of hydrostatic pressure (i.e. lattice isotropic variation) by DP atomic relaxation at absolute zero, which will eliminate the contribution of kinetic energy induced by thermal perturbation. We emphasize that such calculations for amorphous phase are far beyond the capacity of DFT because of calculating spatial scale limitation. In the calculations, the amorphous phase can be stabilized strain-free with the optimized lattice constant 9.005 Å, much smaller than the 9.224 Å for cubic phase, and the corresponding atomic structures are shown in

Supplementary Fig. S8a, b. The resulting energy-versus-lattice diagram is shown in Fig. 6b. It shows that the amorphous phase is energetically favored for lattice constant smaller than ~9.13 Å. As lattice constant is increased, the amorphous phase rapidly becomes unstable and transitions to a "defective" crystalline structure, where a small number of O atoms could not return to its perfect lattice site, forming edge-shared polyhedral defects in cubic phases as shown in Fig. Supplementary S8c (it will be explained in detail below). The energy of these defective cubic phases is very close to that of perfect cubic phase, as shown by green triangles in Fig. 6b. Our results show that the driving force of PIA in $ZrW_2O_8$ is the potential energy.

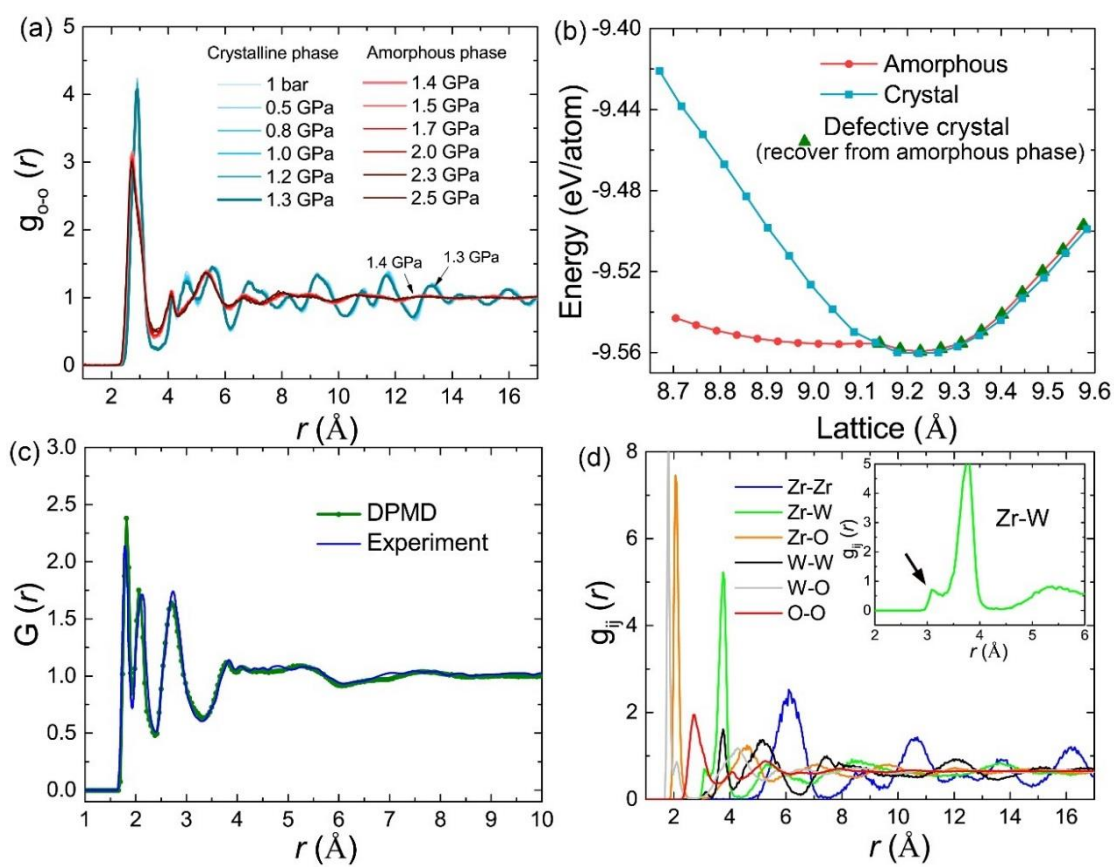

**Fig. 6 Analysis of the atomic structure and potential energy of $ZrW_2O_8$ at different pressures.** (a) Partial O-O RDFs, $g_{O-O}(r)$, of $ZrW_2O_8$ at different pressures from DPMD simulations at 300 K. (b) Energy variations of cubic and amorphous phase $ZrW_2O_8$ with hydrostatic pressure (lattice parameter variation), and amorphous is energy favored for the lattice constant is smaller than ~9.13 Å. (c) Neutron weighted pair distribution functions G(r) and (d) radial distribution function $g_{ij}(r)$ of amorphous $ZrW_2O_8$ at 3 GPa calculated from DP model and experiment.

We proceed to analyze the detailed local structure of the amorphous phase. The DP calculated neutron weighted pair distribution functions G($r$) and partial RDF $g_{ij}(r)$ of the amorphous phase at 300 K and at 3 GPa are shown in Fig. 6c, d. The calculated G(*r*) is almost on top of the experimental data[23]. We analyze the dynamic behavior of atoms during the amorphization transition under pressure. The local dynamic processes are shown in Fig. 7a-c. As mentioned above, $ZrW_2O_8$ crystallizes in cubic phase with the networks of corner-linked octahedral $ZrO_6$ and tetrahedral $WO_4$, which is highly flexible. As the high pressure (above 1.4 GPa) is applied, the one nonbridging O in $W^IO_4$ (green spheres in Fig. 7a) moves into a neighboring octahedral $ZrO_6$, and another one in $W^{II}O_4$ moves to a neighboring tetrahedral $W^IO_4$ (Fig. 7b), forming an edge-shared decahedral $ZrO_7$ and pentahedral $W^IO_6$ structure as shown in Fig. 7c. This edge-shared structure produces the first peak in 3.14 Å of partial RDF of Zr-W (marked by arrows in inset of Fig. 6d). The weakness of that peak intensity indicates that the above dynamic behavior happens randomly throughout the system. The equilibrated snapshot of the supercell after amorphization at 3 GPa is shown in Supplementary Fig. S9, several unique configurations of corner-shared and edge-shared polyhedron were formed by movement of nonbridging O atoms. Examination of the species-specific structure in Fig. 7d (with the corresponding RDFs in Fig. 6d) reveals an interesting aspect of the amorphous phase: the amorphous-like O sit within an almost periodic crystal-like Zr and W coordination array. We note that this structure is in consistent with the proposed Reverse Monte Carlo model of amorphous $ZrW_2O_8$ fitting from experiment data[23].

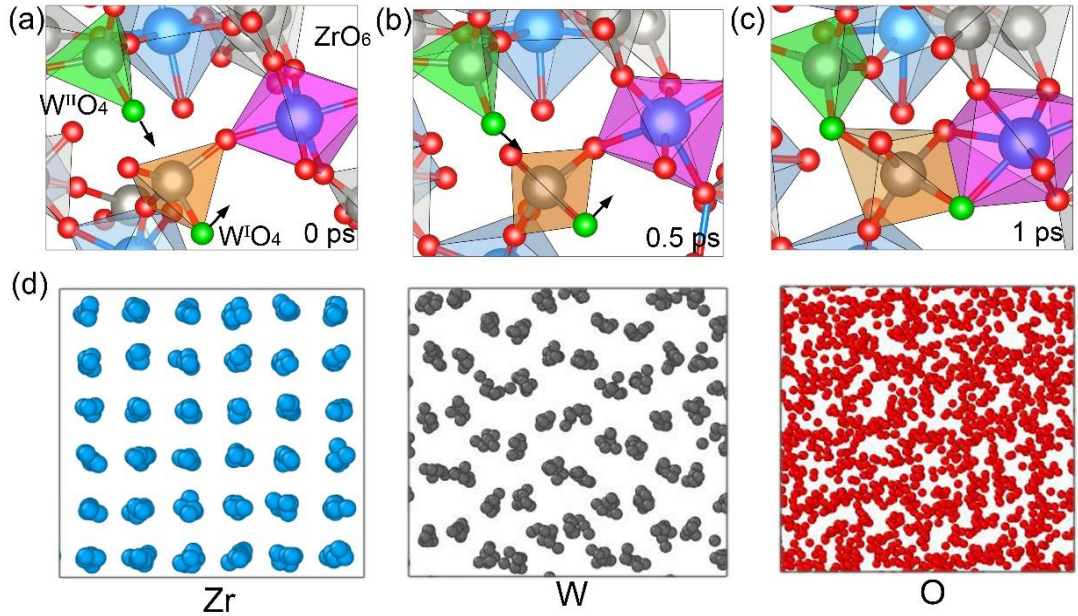

**Fig. 7 Atomic structure evolution of amorphous ZrW$_2$O$_8$ at 3 GPa.** (a)-(c) the movement of nonbridging O (marked in green) result in formation of ZrO$_7$ decanedron and WO$_5$ octahedron. (d) A part snapshot of amorphous 22528-atom supercell, the arrangements of Zr, W, and O are presented respectively.

We next consider the reversibility of the amorphous phase upon decompression. Experiments have confirmed that the PIA in ZrW$_2$O$_8$ is irreversible at room temperature, and recrystallization only occurs when heated above 900 K[2, 40]. In this study, the subsequent DPMD simulations were performed for decompressions from various pressures (0.2 GPa ~ 4.4 GPa) to ambient conditions. The decompression structural property and RDFs results are show in Fig. 5 (the atomic structures are shown in Supplementary Fig. S10) and Fig. 8 (also see Supplementary Fig. S11), and it indicates that pressure-induced amorphous phase can be divided into two categories. (i) For the amorphous phase with 1.4 GPa to 3.7 GPa, significant modifications in the RDF of W-W are observed upon decompression. The second peak at 4.1 Å becomes weaker while a new peak at slightly longer distance, 4.4 Å, appears, which indicates it can recover to crystalline phase after pressure release, so that we defined it as the $a^I$ reversible phase. (ii) For the amorphous with pressure above 3.8 GPa, the RDFs results show that the amorphous phase can be retained upon decompression, which we defined as the $a^{II}$ reversible phase. The same profiles of RDFs of $a^I$ and $a^{II}$ phases in Fig. 8a and

Supplementary Fig. S11a indicate that they have the same equilibrated atomic structure, and they cannot be distinguished by structural characterization in experiment. Therefore, we suggest that amorphous phase undergoes a "hidden" phase transition from $a^I$ to $a^{II}$ amorphous phase at the second critical pressure of 3.8 GPa. The RDF of retained $a^{III}$ amorphous is slightly different from the $a^I$ and $a^{II}$ phase, but it exhibits the almost key features of amorphous phase (Fig. 8b and Supplementary Fig. S11b). The retained $a^{III}$ amorphous phase is quite stable, because it was maintained in subsequent annealing for a long time of 2 ns, as well as at higher temperature of 400 K (Supplementary Fig. S12). It should be emphasized that the critical pressure of "hidden" phase transition slightly depends on compression time. For instance, the compression time increase from 50 ps to 300 ps, the critical pressure decreases from 3.8 GPa to 3.5 GPa (Supplementary Fig. S13).

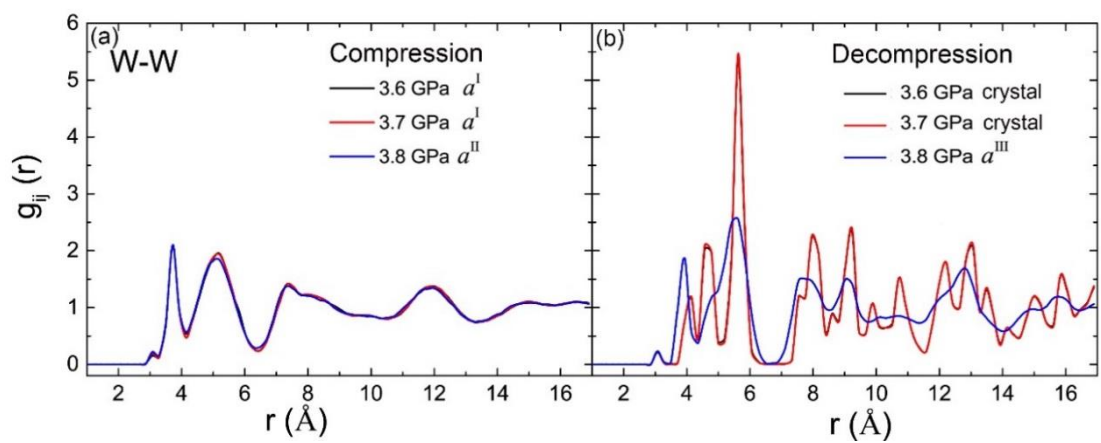

**Fig. 8 Partial W-W RDFs of $ZrW_2O_8$ at different pressures.** (a) RDFs at different pressure, (b) RDFs upon decompression.

For the defective crystalline phase that recovered from $a^I$ phase, the partial W-W RDF is consistent with most of the characteristics of prefect crystalline phase by comparing Supplementary Fig. S11b and Fig. S14, except for an additional $g(r)_{w-w}$ peak in the ~3.16 Å (marked by arrows in Fig. 9a). The first additional of $g(r)_{w-w}$ originates from the edge-shared $WO_5/WO_6$ polyhedrons structures in crystalline phase upon decompression. This edge-shared structure can be regarded as a defect in the crystal, that formed when compression. By comparing the sliced atomic structure of $a^I$ amorphous phase (3.7 GPa) and recovered crystalline phase in Fig. 9c, we found that edge-shared $WO_5/WO_6$ polyhedrons are irreversible during crystallization, which is accord with the RDF result in Fig. 9a. On the contrary, the Fig 9b shows the first peak of $g(r)_{Zr-W}$ disappears in the recovered crystalline phase, which indicates the edge-shared decahedral $ZrO_7$ and octahedral $WO_5$ structure recovers to original octahedral $ZrO_6$ and tetrahedral $WO_4$ upon decompression (atomic structure evolution is shown in Fig. 9d). The entire sliced atomic structures are shown in Supplementary Fig. S15. To sum up, the edge-shared $WO_5/WO_6$ polyhedrons are stable and hard to be destroyed, whereas the edge-shared $ZrO_7$ and $WO_5$ polyhedrons are relatively unstable.

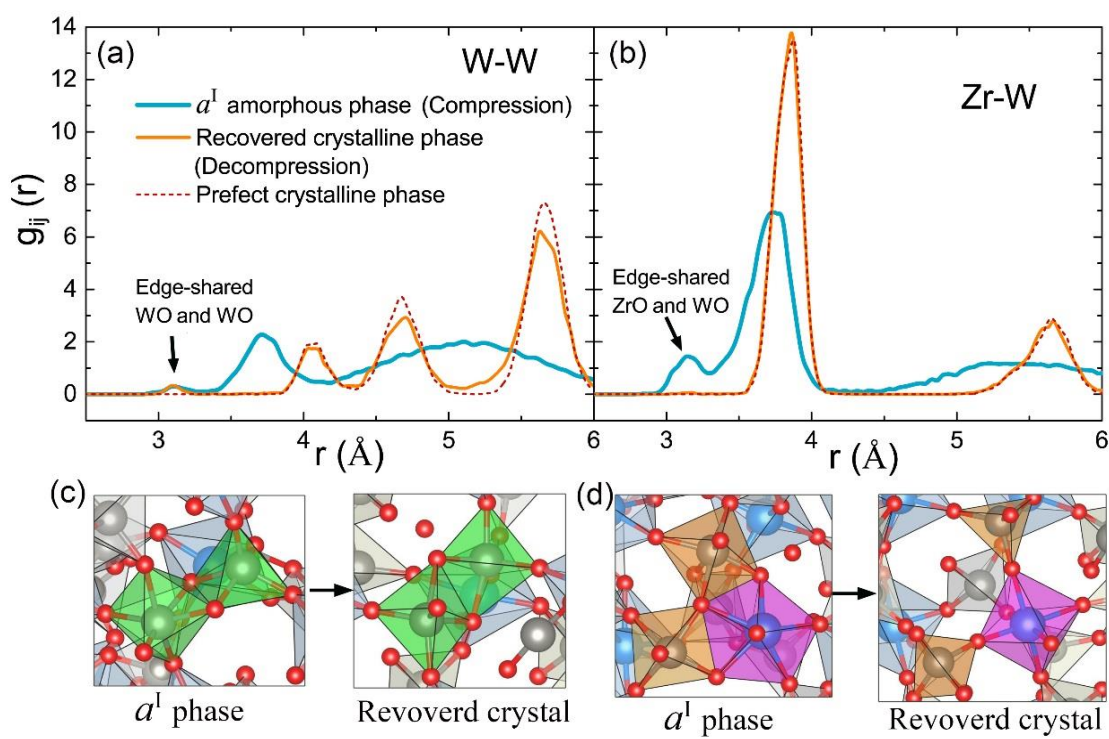

**Fig. 9 Analysis of the atomic structure upon decompression**. The partial RDFs $g(r)_{ij}$ of pressure-induced $a^I$ amorphous (3.7 GPa) and recovered crystalline phase upon decompression for (a) W-W and (b) Zr-W. The first peak of $g(r)_{W-W}$ and $g(r)_{Zr-W}$ marked in figure corresponding to the edge-shared $WO_4/WO_5$ and $ZrO_7$-$WO_6$ polyhedrons. The atomic structural evolution of (c) edge-shared $WO_4/WO_5$ and (d) $ZrO_7$-$WO_6$ polyhedrons from $a^I$ phase to recovered crystalline phase.

## Conclusions

We have demonstrated that ML-based DP can reach an unprecedented level of accuracy in the atomic structural modeling of NTE and PIA materials over a wide temperature and pressure range using the example of $ZrW_2O_8$. The DP calculated NTE behavior and coefficient of thermal expansion in $ZrW_2O_8$ show great agreement with experimental results. By lattice dynamics study of local lattice vibrations, we have shown that both the current RUM and Tent models are only partially reasonable, and the NTE in $ZrW_2O_8$ originate from the large transverse vibrations of the O atoms in the middle of the Zr-O-W linkage. We further investigated the mechanism of PIA in $ZrW_2O_8$. We found that $ZrW_2O_8$ undergoes a first-order potential-driven amorphization transition above a critical pressure of 1.4 GPa, which is well comparable to experimental value. We also reveal that the migration of O atoms leads to additional bond formation that lowers the potential energy, suggesting that the PIA is a potential-driven transition. Most importantly, we predicted that the amorphous phase undergoes a "hidden" phase transition from reversible $a^I$ to irreversible $a^{II}$ phase at a second critical pressure of 3.8 GPa, which shall be verified in future experiments. Our study on $ZrW_2O_8$ not only have important implications for future research on other NTE and PIA materials, but also opens the door for quantitatively accurate atomistic modeling of general materials with direct links to experiments. We believe that the DP method will play a pivotal role in future design of structural and functional materials whose properties depend sensitively on detailed atomic configurations.

## Methods

**1. Concurrent learning procedure**

Carefully choose the $ZrW_2O_8$ configurations that used in the training dataset is crucial to the success of training process and accurate of DP. Because we focus on the NTE and PIA behaviors of $ZrW_2O_8$, we need all the characteristic configurations of crystal and amorphous at a wide temperature and pressure range. Most of configurations that we need are far from the ground state of cubic phase. In this work, we use the DP Generator (DP-Gen) to generate a set of training data that covers an enough wide range of relevant configurational space efficiently. DP-Gen is a concurrent learning strategy[36], and the workflow of each iteration includes three main steps as illustrated in Fig. 10: (1) Training the deep learning potentials, (2) exploration configurations by DPMD simulations, (3) labeling configurations according to certain criterion and added into training dataset.

We start with DFT optimized ground-state structures of $P2_13$ cubic phase $ZrW_2O_8$. The DFT calculated lattice constant with Perdew-Burke-Ernzerhof solid (PBEsol) is 9.215 Å, slightly larger than 9.160 Å in experiment[11]. In first iteration of DP-Gen workflow, the configurations of initial training dataset contain 200 randomly perturbed structures of 44-atoms cubic cell. The maximum magnitude of perturbed displacement for atomic coordinates is 0.01 Å, and the strain is −0.003 to 0.003 of the ground-state lattice parameters. Starting with above training datasets, four different DPs are trained using DeepMD method[26, 27], based on different values of deep neural network parameters. The exploration step was performed in which one of the DPs is used for MD simulations at given pressure and temperature to explore the configuration space. For all sampled configurations in MD trajectories, the other three DPs will predict the atomic forces of all atoms. The maximum deviation of the four DPs predicted forces ($\sigma_f^{max}$) can be used to formulate the criterion for labeling configurations:

$$\sigma_f^{max} = \max\sqrt{\langle |F_i - \langle F_i \rangle|^2 \rangle},$$

Where $\langle ... \rangle$ indicates the average of DP predicted force, and $F_i$ denotes the predicted force on the atom *i*. The explored configurations possess the characteristic of $\sigma_f^{max} < \sigma_{low}$ indicates they are already well described with a high accuracy by the current DP, which are labeled as accuracy, whereas a explored configuration with $\sigma_f^{max} > \sigma_{high}$ is highly distorted and unphysical due to poor model quality of DP and is thus labeled as failure. The failure ratio is often high in the first several iterations. Only configurations satisfying $\sigma_{low} < \sigma_f^{max} < \sigma_{high}$ are selected as candidates for further self-consistent DFT calculation and are added to the training dataset for training in the next iteration. The $\sigma_f^{max}$ can be also used as the convergence criterion for DP-Gen iterations, and the DP-Gen iterations are considered converged when the accuracy ratio is large than 99 %. Here, the $\sigma_{low}$ and $\sigma_{high}$ are set to 0.05 and 0.15 eV/Å in the iterations for 50 K to 550 K, and 0.08 and 0.25 eV/Å in the iterations for 600 K to 1500 K to ensure the convergence. Finally, by iterating above procedure 56 times, 20460 configurations were obtained for training. For more details of the DP-Gen process, please refer to the original literature or our previous work of developing SrTiO$_3$ DP[31, 36].

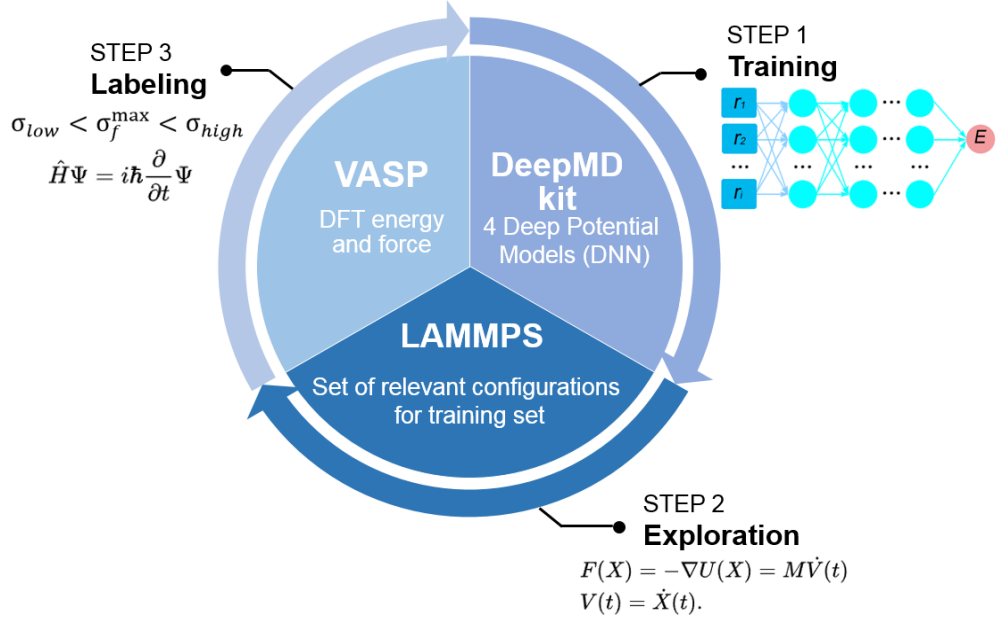

**Fig. 10 The deep learning potential training procedure, including training, exploration, and labeling.**

## 2. Training the DP

The DeePMD-kit code is used for training of DP[41]. In the deep neural network of DP, the total potential energy of a configuration is assumed to be a sum of each atomic energy, which is mapped from a descriptor through an embedding network. The descriptor characterizes the local environment of atom within a cutoff radius $R_c$. Here, the $R_c$ is set to 6 Å. The maximum number of atoms within the $R_c$ is set to 50 for Zr, 100 for W, and 400 for O, respectively. The translational, rotational, and permutational symmetry of the descriptor are preserved by another embedding network. The smooth edition of the DP model was employed to remove the discontinuity introduced by the $R_c$[27]. The sizes of the embedding and fitting networks are (25, 50, 100) and (240, 240, 240), respectively. The loss function that used in this work have the same form as our recent work[31]. The weight coefficients of the energy, atomic force and virial terms in the loss functions change during the optimization process from 0.2 to 1, 500 to 1, and 0.02 to 0.2. The DPs are trained with 1 800 000 steps by learning rate decaying from 1e−3 to 3.5e−8 exponentially. The model compression scheme was applied in this work for boosting the computational efficiency of the DPMD simulation[42].

3. **DFT calculations**

The initial training dataset in the first iteration is obtained by performing a 10-step *ab initio* MD simulation for randomly perturbed 44 atoms cubic cell at 50 K. After labeling candidate configurations, self-consistent DFT calculations were performed subsequently. The DFT calculations were performed using a plane-wave basis set with a cutoff energy of 500 eV as implemented in the Vienna Ab initio Simulation Package [43, 44], and the electron exchange-correlation potential was described using the generalized gradient approximation and PBEsol scheme[45]. The Brillouin zone was sampled with a 4 × 4 × 4 Monkhorst-Pack k-point grid for the cubic unit cell.

4. **Molecular dynamics simulations**

In exploration step, the MD simulations were performed using LAMMPS code with periodic boundary conditions[46]. The MD simulations adopt the isobaric-isothermal (NPT) ensemble with temperature set from 10 to 1500 K and pressure set from 1 bar to 5 GPa, because the explorative temperature and pressure should be large than the ones for subsequent NTE and PIA simulations. A Nose-Hoover thermostat and Parrinello-Rahman barostat are employed to control temperature and pressure, respectively[47, 48]. The time step in simulations is set to 1 fs.

The optimized DP can be used to study the atomic dynamics driven by temperature and pressure via performing deep potential MD (DPMD) simulations. The DPMD simulations starting with 8 × 8 × 8 supercell (i.e. 7.3 nm ×7.3 nm ×7.3 nm) of optimized cubic α-$ZrW_2O_8$ phase containing 22528 atoms. All results about NTE and PIA of $ZrW_2O_8$ were executed by DPMD simulations. For NTE simulations, the equilibrium run is 50 ps, followed by a production run of 100 ps at a specified temperature. The lattice constant, true/apparent bond length (angle) were calculated by the average of

200 snapshots in 100ps of equilibrium state. For PIA simulations, cubic symmetry was imposed (isotropic changes in cell lengths) during NPT simulations. The "lattice constants" of pressure-induced amorphous phase is defined as $\sqrt[3]{V}/8$, where V is the volume of the simulated 8 × 8 × 8 supercell. The simulation time for compression and decompression were set to 50 ps and 100 ps, respectively. To prove the stability of retained $a^{III}$ amorphous phase, we lengthen the decompression time to 2 ns, and found the $a^{III}$ amorphous phase is maintained (Supplementary Fig. S11). We also lengthen the compression time to prove the second critical pressure of 3.8 GPa slightly depends on compression time (Supplementary Fig. S12).

## Data availability

All the input files, final training datasets, and DP model files to reproduce the results contained in this paper are available in DP-library website https://dplibrary.deepmd.net/ [49].

## Code availability

The DFT calculations, MD simulations, concurrent learning procedure, and deep neural network training were performed with VAPS, LAMMPS, DP-GEN, and DeePMD-kit code. The latter three are open source codes available at https://lammps.sandia.gov and https://github.com/deepmodeling.

## Acknowledgements

This paper was supported by the National Key R&D Program of China (Grants No. 2021YFA0718900, and No. 2017YFA0303602), the Key Research Program of Frontier Sciences of CAS (Grant No. ZDBS-LY-SLH008), the National Nature Science Foundation of China (Grants No. 11974365), the K.C. Wong Education Foundation (GJTD-2020-11).

## Additional information

Supplementary information is available for this paper at https://doi.org/XXXX